\documentclass[11pt]{article}
\usepackage{amssymb}

\newcommand{\reseteqnum}{\setcounter{equation}{0}}

\newcommand{\plb}[3]{Phys. Lett. {\bf B#1} (#2) #3} 
\newcommand{\prl}[3]{Phys. Rev. Lett. {\bf #1} (#2) #3}
\newcommand{\prd}[3]{Phys. Rev. {\bf D#1} (#2) #3}
\newcommand{\npb}[3]{Nucl. Phys. {\bf B#1} (#2) #3}

\title{
\hfill
\parbox{3cm}{\normalsize DPNU-99-37\\
{\tt  hep-lat/9912056}}\\
\vspace{0.5cm}
Locality bound for effective four-dimensional action \\
of domain-wall fermion
\author{
Yoshio Kikukawa\thanks{e-mail address:
kikukawa@eken.phys.nagoya-u.ac.jp} 
\\
{\normalsize\em Department of Physics, Nagoya University 
}\\
{\normalsize\em Nagoya 464-8602, Japan}
\\
\\
\date{\normalsize December, 1999}
}
}
\begin{document}

\maketitle

\begin{abstract}

We discuss locality in the domain-wall QCD
through the effective four-dimensional Dirac operator 
which is defined by the transfer matrix of the five-dimensional 
Wilson fermion. We first derive an integral representation for 
the effective operator, using the inverse five-dimensional
Wilson-Dirac operator with the anti-periodic boundary condition 
in the fifth direction. Exponential bounds are obtained from it 
for gauge fields with small lattice field strength.

\end{abstract}

\newpage

\section{Introduction}
\label{sec:introduction}
\reseteqnum

Locality properties of Neuberger's lattice Dirac 
operator \cite{overlap-D}, 
which is derived from the overlap formalism \cite{overlap} and 
satisfies the Ginsparg-Wilson relation \cite{ginsparg-wilson-rel}, 
has been examined by 
Hern\'andes, Jansen and L\"uscher \cite{locality-of-overlap-D}.
For a certain class of gauge fields with small lattice field strength,
exponential bounds have been proved rigorously on the kernels of 
the Dirac operator and 
its differentiations with respect to the gauge field.
These properties assure that the index theorem holds true
on the lattice \cite{index-theorem-at-finite-lattice,overlap}. 
The index theorem implies the topological properties 
of chiral anomalies. It plays the crucial role in the recent
construction of lattice chiral gauge theories \cite{chiral-anomaly-abelian,
abelian-chiral,nonabelian-chiral,fujiwara-suzuki-wu,adams-6d,lat99-luscher}.
Numerical studies of the locality of Neuberger's lattice Dirac 
operator are also found in \cite{locality-of-overlap-D,borici}.

The purpose of this paper is to argue the locality 
of the low energy effective action 
of the domain-wall fermion \cite{domain-wall-fermion,boundary-fermion,
boundary-fermion-QCD} 
\cite{blum-soni,blum-soni-wingate}
\cite{vranas-pauli-villars}
\cite{DWF-columbia}
\cite{DWF-tsukuba-oneloop-renormalization-factor,
DWF-tsukuba-scaling,DWF-tsukuba-fine-coarse}
\cite{epsilon-prime-over-epsilon,DWF-RBC}
\cite{chiu-DWF}.
It has been known that the partition function of the domain-wall fermion,
in the anti-periodic subtraction scheme \cite{vranas-pauli-villars},
reduces to a single determinant of an effective four-dimensional 
Dirac operator \cite{truncated-overlap},
\begin{equation}
\label{eq:effective-Dirac-operator-finiteN}
D_{\rm eff}^{(N)} = \frac{1}{2a}\left(1 + \gamma_5 \tanh \frac{N}{2} a_5 
\widetilde H \right) .
\end{equation}
Here $N$ and $a_5$ denote the lattice size and the lattice spacing 
of the fifth dimension, respectively. $\widetilde H$ is defined through 
the transfer matrix of the five-dimensional Wilson fermion
\begin{equation}
\label{eq:transfer-matrix}
T = e^{- a_5 \tilde H} 
= \left(
\begin{array}{cc} \frac{1}{B} & - \frac{1}{B} C \\
                 -C^\dagger \frac{1}{B} 
& B + C^\dagger \frac{1}{B} C 
\end{array} \right),
\end{equation}
%
where
\footnote{
In this expression,
the positivity of $B$
is required for the transfer matrix to be defined consistently.
It is assured when $ 0 < \frac{a_5}{a} m_0  < 1$ . 
It is also assumed that $N$ is even.
}
\begin{eqnarray}
\label{eq:operator-C}
  C &=& a_5 \, \sigma_\mu \, 
\frac{1}{2}\left(\nabla_\mu+\nabla_\mu^\ast\right) , \\
\label{eq:operator-B}
 B &=& 1 + a_5 \left( 
-\frac{a}{2} \nabla_\mu\nabla_\mu^\ast - \frac{m_0}{a} 
\right) .
\end{eqnarray}
%
%
The limit $N\rightarrow \infty$ is defined well
as long as $\widetilde H^2 > 0 $.
The effective Dirac operator 
Eq.~(\ref{eq:effective-Dirac-operator-finiteN}) then
reduces to Neuberger's lattice Dirac operator using $\widetilde H$,
\begin{equation}
\label{eq:effective-Dirac-operator}
D_{\rm eff} 
= 
\frac{1}{2a}
\left(1 + \gamma_5 \frac{\widetilde H}{\sqrt{\widetilde H^2}}\right),
\end{equation}
and turns out to satisfy the Ginsparg-Wilson relation.

Moreover, the propagator of the light fermion field
which is introduced by Furman and 
Shamir \cite{boundary-fermion-QCD},
\begin{equation}
q(x)= \psi_L(x,1) + \psi_R(x,N), \qquad
\bar q(x)= \bar \psi_L(x,1) + \bar \psi_R(x,N) ,
\end{equation}
can be expressed in terms 
of the effective Dirac operator \cite{kikukawa-noguchi}:
\begin{equation}
\label{eq:light-fermion-propagator}
\left\langle q(x) \bar q(y) \right\rangle 
= \frac{a_5}{a^4} 
\left( \frac{1}{a} { D_{\rm eff}^{(N)}}^{-1}-\delta(x,y) \right) .
\end{equation}
The anomalous term in the axial Ward-Takahashi identity
\begin{equation}
\label{eq:anomalous-chiral-symmetry-breaking-WTI}
X^{(N)}(x) = 
\frac{2}{a_5} \left\{
 \bar \psi_L(x,\frac{N}{2}+1) \psi_R(x,\frac{N}{2}) 
-\bar \psi_R(x,\frac{N}{2}) \psi_L(x,\frac{N}{2}+1) 
\right\} 
\end{equation}
can also be expressed with it:
\begin{eqnarray}
\label{eq:anomalous-chiral-symmetry-breaking}
a^4 \left\langle X^{(N)}(x) \right\rangle &=& 
  {\rm tr}\gamma_5 \, 2\left( 1- a D_{\rm eff}^{(N)} \right) (x,x)
\nonumber\\
&& 
- {\rm tr} \left(\frac{1}{a} { D_{\rm eff}^{(N)}}^{-1} \gamma_5 
  \frac{1}{\cosh^2 \frac{N}{2}a_5 \widetilde H } \right) (x,x) .
\end{eqnarray}
In the limit $N\rightarrow \infty$, 
this reduces to the chiral anomaly associated with the exact chiral 
symmetry \cite{exact-chiral-symmetry,
kikukawa-yamada,adams,suzuki,fujikawa,chiu},
\begin{equation}
a^4 \left\langle X (x) \right\rangle=
{\rm tr}\gamma_5 \, 2\left( 1- a D_{\rm eff} \right) (x,x) .
\end{equation}
It would have the topological properties, 
if the effective Dirac operator is local and depends 
smoothly on the gauge fields.

In view of this direct relation,\footnote{
In Eq.~(\ref{eq:light-fermion-propagator})
the propagator of the boundary variables turns out to 
be chiral invariant in the limit $N \rightarrow \infty$ 
due to the negative contact term. 
In this respect, it is important to note the contribution of 
massive modes (including the Pauli-Villars modes),
which takes account of the chiral anomaly in 
Eqs.~(\ref{eq:anomalous-chiral-symmetry-breaking-WTI})
and (\ref{eq:anomalous-chiral-symmetry-breaking})
and fills the gap to the Ginsparg-Wilson fermion \cite{kikukawa-noguchi}.
Following \cite{exact-chiral-symmetry,luscher-private,rational-approximation},
we may also introduce an ``heavy'' auxiliary field 
$  a^4 \sum_x \bar \chi(x) \chi(x) $ 
and redefine the boundary variables as
$q^\prime(x) = q(x) +\chi(x)$,
$\bar q^\prime(x) = \bar q(x) +\bar \chi(x)$,
so that the contact term in the propagator is removed.
It is also pointed out in \cite{borici2} that 
a certain modification of the action of the domain-wall fermion 
leads directly to the propagotor of the boundary variables 
which satisfies the Ginsparg-Wilson relation.
} 
it seems reasonable to argue locality in the domain-wall fermion 
approach through the locality properties of the effective Dirac 
operators Eq.~(\ref{eq:effective-Dirac-operator-finiteN}) and 
Eq.~(\ref{eq:effective-Dirac-operator}).
It is expected that a similar exponential bound could be established
under certain conditions, like the result obtained 
by Hern\'andez, Jansen and L\"uscher 
\cite{locality-of-overlap-D}.
In our case, though, the hermitian Wilson-Dirac operator
should be replaced with $\widetilde H$. 
Then the use of the Legendre polynomials \cite{locality-of-overlap-D}
does not lead to the expansion in terms of operators with finite ranges. 

For our purpose, we first derive an integral 
representation for the effective Dirac operator. The inverse square 
root in Neuberger's lattice Dirac operator can be written by the integral:
\begin{equation}
\frac{1}{2} \left(1+ \gamma_5 \frac{H}{\sqrt{H^2}} \right)
= 
\frac{1}{2}+\gamma_5 \int^\infty_{-\infty}\frac{dp}{2\pi} 
\frac{1}{i \gamma_5 p + \left(D_{\rm w}- \frac{m_0}{a}\right)} \gamma_5 .
\end{equation}
Corresponding to this, we can show that 
the effective Dirac operator admits 
the following representation \cite{kikukawa-noguchi}:
\begin{eqnarray}
\label{eq:effective-Dirac-operator-Intro}
a D_{\rm eff}^{(N)} 
&=&  1- 
  P_R \left\{ a_5 \overline{D}_{\rm 5w} \right\}^{-1}_{NN} P_L 
- P_L \left\{ a_5 \overline{D}_{\rm 5w} \right\}^{-1}_{11} P_R 
\nonumber\\
&& \quad 
- P_R \left\{ a_5 \overline{D}_{\rm 5w} \right\}^{-1}_{N1} P_R
- P_L \left\{ a_5 \overline{D}_{\rm 5w} \right\}^{-1}_{1N} P_L ,
\nonumber\\
\end{eqnarray}
where $\overline{D}_{\rm 5w}$ is  the five-dimensional 
Wilson-Dirac operator {\em with the anti-periodic boundary condition}
in the fifth-dimension. Its inverse may be expressed as
\begin{equation}
\left\{a_5  \overline{D}_{\rm 5w}  \right\}^{-1}_{st}  
= \frac{1}{N}\sum_p 
\frac{e^{i p(s-t)}}
{i \gamma_5 \sin p+ 1-\cos p
        + a_5 \left(D_{\rm w}- \frac{m_0}{a}\right) }.
\end{equation}
The summation is taken over the discrete 
momenta ${p=\frac{2\pi}{N}(k-\frac{1}{2})}$ $(k=1,2,\cdots,N)$ and
in the limit $N\rightarrow\infty$ it reduces to the continuous integral.

From this representation, it is rather clear that the effective 
Dirac operator can be defined consistently if the five-dimensional 
Wilson-Dirac operator with the anti-periodic boundary condition is 
not singular and invertible for all $N$.
In this respect, we should note that 
the lower bound on the square of the five-dimensional 
Wilson-Dirac operator is related closely 
to that on the square of the four-dimensional 
Wilson-Dirac operator \cite{locality-of-overlap-D,bound-neuberger},
because the gauge field is four-dimensional.
In fact, the same lower bound can be set for the class 
of gauge fields with small lattice field strength. 

Given the positive lower bound on the square of the five-dimensional 
Wilson-Dirac operator, it is possible to 
formulate a series expansion in terms of the five-dimensional
Wilson-Dirac operator, using the generating function 
of the Chebycheff polynomials \cite{bunk}.
The exponential bounds on the effective Dirac operator and its
differentiations can be established from it. 

We may also discuss the Ginsparg-Wilson relation of 
the effective Dirac operator through this integral representation. 
We will see that this reduces to the question concerning
the property of the five-dimensional Dirac operator 
under the chiral transformation introduced by Furman and 
Shamir \cite{boundary-fermion-QCD}.

Another interesting aspect of the effective Dirac operator 
is its behavior in case with the singular gauge configuration 
for which isolated eigenvalues of the hermitian Wilson-Dirac operator
collapse to zero. In \cite{locality-of-overlap-D}, it has been 
proved rigorously that Neuberger's Dirac operator in terms
of the hermitian Wilson-Dirac operator remains local even with
such singular gauge configurations. We will argue that
it is also true for the effective Dirac operator 
Eq.~(\ref{eq:effective-Dirac-operator}).

The approach to the chiral symmetry limit from a finite $N$ 
is the most important issue for the practical implementation 
of exact chiral symmetry using the domain-wall fermion 
\cite{Edwards-et-al-small-eigenvalues-localization,
Edwards-et-al-small-eigenvalues-distribution}
\cite{DWF-columbia,DWF-tsukuba-fine-coarse,DWF-RBC}
\cite{shamir-modified-PF,DWF-strong-coupling-expansion,better-DWF}.
In this respect, our result of the locality and exact chiral symmetry 
of the effective four-dimensional action 
is restricted for the class of gauge fields with small lattice field strength.
It is a nonperturbative result and it gives a sufficient condition
for that the exact chiral symmetry based on the Ginsparg-Wilson 
relation can be implemented using the domain-wall fermion.
But it does not assure that it would work practically
in the numerical simulations using the standard Wilson's gauge action. 
Our result, however, presents an explicit method to connect
the locality and chiral symmetry properties of the domain-wall fermion 
to the spectrum of the four-dimensional Wilson-Dirac operator.
We hope that such a method would be useful in order to study the above 
practical issue.

This paper is organized as follows. 
In section~\ref{sec:domain-wall-fermion}, we briefly review the
domain-wall fermion in order to fix our notation. 
In section~\ref{sec:Effective-Dirac-operator-Integral-rep},
we describe how to derive the integral representation for the 
effective Dirac operator. 
We also discuss how the Ginsparg-Wilson relation for the 
effective Dirac operator follows in this integral representation.
In section~\ref{sec:positivity}, we discuss the positivity
of the five-dimensional Wilson-Dirac operator with
the anti-periodic boundary condition.
With this result, we consider exponential bounds 
for the effective Dirac operator 
and its differentiations in section~\ref{sec:exponential-bound}.
We also discuss the locality in the case with the singular gauge 
configurations. 
In section~\ref{sec:summary-discussion}, we summarize our result
and give some discussions concerning the issue of the 
approach to the chiral symmetry limit.

\section{Domain-wall fermion in the anti-periodic subtraction scheme}
\label{sec:domain-wall-fermion}
\reseteqnum

In this section, we review the domain-wall 
fermion \cite{domain-wall-fermion, boundary-fermion,boundary-fermion-QCD}
and fix our notation.
The domain-wall fermion is defined by the five-dimensional
Wilson-Dirac fermion with the Dirichlet boundary condition.
\begin{equation}
S_{\rm DW} = \sum_{t=1}^N a^4 \sum_x \bar \psi(x,t) D_{\rm 5w} \psi(x,t),
\end{equation}
\begin{equation}
D_{\rm 5w} =   
 \gamma_\mu \frac{1}{2}\left(\nabla_\mu+\nabla_\mu^\ast\right) \delta_{st} 
+ P_L M_{st} + P_R M^\dagger_{st} .
\end{equation}
We assume the lattice size of the fifth dimension $N$ is even.
For $N=6$, the mass matrix reads
\begin{eqnarray}
\label{eq:mass-matrix}
M_{st}&=&
\frac{1}{a_5}
\left( \begin{array}{cccccc}
              B & -1 & 0 & 0 & 0 & 0 \\
              0 & B & -1 & 0 & 0 & 0 \\
              0 & 0 & B & -1 & 0 & 0 \\
              0 & 0 & 0 & B & -1 & 0 \\
              0 & 0 & 0 & 0 & B  & -1 \\
              0 & 0 & 0 & 0 & 0 & B  
\end{array} \right),
\end{eqnarray}
where $B$ is defined by 
\begin{equation}
\label{eq:operator-B}
 B = 1 + a_5 \left( 
-\frac{a}{2} \nabla_\mu\nabla_\mu^\ast - \frac{m_0}{a} 
\right) .
\end{equation}
The chiral transformation is introduced as vector-like one so that
the symmetry breaking is minimized \cite{boundary-fermion-QCD}:
\begin{eqnarray}
  \delta \psi(x,t)&=& - \psi(x,t)  \qquad  t \le \frac{N}{2} , \\
  \delta \psi(x,t)&=& + \psi(x,t)  \qquad  t \ge \frac{N}{2}+1 .
\end{eqnarray}
Accordingly, the anomalous term is given by
\begin{equation}
X^{(N)}(x) = 
\frac{2}{a_5} \left\{
 \bar \psi_L(x,\frac{N}{2}+1) \psi_R(x,\frac{N}{2}) 
-\bar \psi_R(x,\frac{N}{2}) \psi_L(x,\frac{N}{2}+1) 
\right\} 
\end{equation}

The partition function of the domain-wall fermion may be defined with
the subtraction of the Pauli-Villars fields, which is subject to 
the anti-periodic boundary condition in the fifth
dimension. \footnote{For this subtraction to work consistently, 
we should require the positivity of 
$\overline{D}_{\rm 5w}$:
\begin{equation}
   \overline{D}_{\rm 5w}^\dagger \overline{D}_{\rm 5w}   > 0 .
\end{equation}
We will see later that this requirement also assures the locality
and the Ginsparg-Wilson relation of the effective Dirac operator. 
}
\begin{equation}
Z_{\rm DW}= \frac{ \det D_{\rm 5w} }
                            { \det \overline{D}_{\rm 5w}} ,
\end{equation}
where
\begin{equation}
\overline{D}_{\rm 5w} =   
 \gamma_\mu \frac{1}{2}\left(\nabla_\mu+\nabla_\mu^\ast\right) \delta_{st} 
+ P_L \overline{M}_{st} + P_R \overline{M}^\dagger_{st} ,
\end{equation}
\begin{eqnarray}
\label{eq:mass-matrix-PV}
\overline{M}_{st}&=&
\frac{1}{a_5}
\left( \begin{array}{cccccc}
              B & -1 & 0 & 0 & 0 & 0 \\
              0 & B & -1 & 0 & 0 & 0 \\
              0 & 0 & B & -1 & 0 & 0 \\
              0 & 0 & 0 & B & -1 & 0 \\
              0 & 0 & 0 & 0 & B  & -1 \\
              1 & 0 & 0 & 0 & 0 & B  
\end{array} \right), \qquad (N=6)
\end{eqnarray}

For later convenience, we perform a chirally asymmetric parity 
transformation in the fifth dimension:
\begin{eqnarray}
  \psi(x,t) &=& \left( P_R + P_L P \right)_{ts} \psi^\prime(x,s),  \\
  \bar \psi(x,t) &=& \bar \psi^\prime(x,s) \left( P_R P + P_L \right)_{st} ,
\end{eqnarray}
where
\begin{equation}
  P_{st} = \left( \begin{array}{cccccc}
          0 & 0 & 0 & 0 & 0 & 1 \\
          0 & 0 & 0 & 0 & 1 & 0 \\
          0 & 0 & 0 & 1 & 0 & 0 \\
          0 & 0 & 1 & 0 & 0 & 0 \\
          0 & 1 & 0 & 0 & 0 & 0 \\
          1 & 0 & 0 & 0 & 0 & 0 
 \end{array} \right) \quad (N=6). \\
\end{equation}
Accordingly, the five-dimensional Dirac operators are transformed as 
follows:
\begin{eqnarray}
D_{\rm 5w}^\prime 
&=& \left( P_R P + P_L \right) D_{\rm 5w}
\left( P_R + P_L P \right) \\
&=&   
 \gamma_\mu \frac{1}{2}\left(\nabla_\mu+\nabla_\mu^\ast\right) P_{st} 
+ M^H_{st}  ,\\
\overline{D}_{\rm 5w}^\prime 
&=& \left( P_R P + P_L \right) \bar D_{\rm 5w}
 \left( P_R + P_L P \right) \\
&=&   
 \gamma_\mu \frac{1}{2}\left(\nabla_\mu+\nabla_\mu^\ast\right) P_{st} 
+ \overline{M}_{st}^H ,
\end{eqnarray}
where, for $N=6$,
\begin{eqnarray}
\label{eq:hermitian-mass-matrix}
M^{\rm H}_{st}&=& M_{st} P =P M^\dagger_{st}  \nonumber\\ 
&=&
\frac{1}{a_5}
\left( \begin{array}{cccccc}
           0 & 0 & 0 & 0 & -1& B\\
           0 & 0 & 0 & -1& B & 0 \\
           0 & 0 & -1& B & 0 & 0 \\
           0 &-1 & B & 0 & 0 & 0 \\
           -1& B & 0 & 0 & 0 & 0 \\ 
           B & 0 & 0 & 0 & 0 & 0 
\end{array} \right) . \nonumber\\
\end{eqnarray}
\begin{eqnarray}
\label{eq:hermitian-mass-matrix-antiperiodic}
\overline{M}_{st}&=&
\frac{1}{a_5}
\left( \begin{array}{cccccc}
           0 & 0 & 0 & 0 & -1& B\\
           0 & 0 & 0 & -1& B & 0 \\
           0 & 0 & -1& B & 0 & 0 \\
           0 &-1 & B & 0 & 0 & 0 \\
           -1& B & 0 & 0 & 0 & 0 \\ 
           B & 0 & 0 & 0 & 0 & 1
\end{array} \right) . \nonumber\\
\end{eqnarray}

In this basis, 
the chiral transformation adopted by Shamir and Furman 
\cite{boundary-fermion-QCD}
can be expressed as follows:
\begin{equation}
\label{eq:chiral-transformation-of-DW}
\delta \psi'_s(x) 
= \left( \Gamma_5 \right)_{st} \psi'_t(x),
\end{equation}
where $\Gamma_5$ is given (for $N=6$) by
\begin{equation}
 \left( \Gamma_5 \right)_{st} = 
 \left( \begin{array}{cccccc} -\gamma_5 & 0 & 0 & 0 & 0 & 0 \\
                              0 & -\gamma_5 & 0 & 0 & 0 & 0 \\
                              0 & 0 & -\gamma_5 & 0 & 0 & 0 \\
                              0 & 0 & 0 & \gamma_5 & 0 & 0 \\
                              0 & 0 & 0 & 0 & \gamma_5 & 0 \\
                              0 & 0 & 0 & 0 & 0 & \gamma_5 
         \end{array} 
 \right)  \qquad (N=6).
\end{equation}
With this definition of the chiral transformation, 
$D^\prime_{\rm 5w}$ and 
$\overline{D}^\prime_{\rm 5w}$ satisfy the following identities,
respectively, 
\begin{eqnarray}
\label{eq:chiral-property-domain-wall-D}
\left\{ \Gamma_5 D_{\rm 5w}^\prime 
       + D_{\rm 5w}^\prime \Gamma_5 \right\}_{st}
&=&\frac{2}{a_5} \, 
 \gamma_5 \delta_{s \frac{N}{2}}\delta_{t \frac{N}{2}} , \\
\label{eq:chiral-property-PV-D}
\left\{ \Gamma_5 \overline{D}_{\rm 5w}^\prime 
       + \overline{D}_{\rm 5w}^\prime \Gamma_5 \right\}_{st}
&=&\frac{2}{a_5} \, 
 \gamma_5 \delta_{s \frac{N}{2}}\delta_{t \frac{N}{2}}
+\frac{2}{a_5} \, 
 \gamma_5 \delta_{s N}\delta_{t N } .
\end{eqnarray}
The chiral symmetry breaking occurs at $t=\frac{N}{2}$ 
in the five-dimensional Dirac operator
for the domain-wall fermion. 
On the other hand, 
it occurs both  at $t=\frac{N}{2}$ and at $t=N$
for the Pauli-Villars field, because of the anti-periodic 
boundary condition. 

\section{Effective four-dimensional Dirac operator}
\label{sec:Effective-Dirac-operator-Integral-rep}
\reseteqnum

\subsection{An integral representation of the effective 
four-dimensional Dirac operator
}
\label{subsec:Integral-rep}
\reseteqnum

The functional determinant of the domain-wall fermion, 
in the anti-periodic subtraction scheme,  
reduces to a single determinant of a four-dimensional Dirac operator,
\begin{equation}
\frac{ \det D_{\rm 5w} }{ \det \overline{D}_{\rm 5w}}
=\det a D_{\rm eff}^{(N)} .
\end{equation}
In this section, we reproduce this result and 
derive an integral representation for 
the effective four-dimensional Dirac operator.

We may write the partition function as follows:
\begin{eqnarray}
\frac{ \det D_{\rm 5w} }{ \det \overline{D}_{\rm 5w}}
&=& \frac{ \det D^\prime_{\rm 5w} }{ \det \overline{D}^\prime_{\rm 5w}}
= \det \left( D^\prime_{\rm 5w} 
              \left\{ \overline{D}^\prime_{\rm 5w} \right\}^{-1}\right) .
\end{eqnarray}
Then, we note a simple relation between two five-dimensional Wilson-Dirac
operators:
\begin{equation}
\label{eq:Dirichlet-Anti-periodic-relation}
D^\prime_{\rm 5w} =   \overline{D}^\prime_{\rm 5w} 
- \frac{1}{a_5}\delta_{sN} \delta_{Nt}.
\end{equation}
This relation implies that 
\begin{equation}
 D^\prime_{\rm 5w} \left\{  \overline{D}^\prime_{\rm 5w} \right\}^{-1}
= \delta_{st} -   
\frac{1}{a_5}
\delta_{sN} \left\{ \overline{D}^\prime_{\rm 5w} \right\}^{-1}_{1t} .
\end{equation}
Since this matrix is lower triangle in the lattice
indices of the fifth dimension, we can easily see that 
its determinant reduces to a single four-dimensional determinant:
\begin{equation}
  \det \left( D^\prime_{\rm 5w} 
              \left\{ \overline{D}^\prime_{\rm 5w} \right\}^{-1}\right) 
= \det \left( 1- \frac{1}{a_5}
\left\{ \overline{D}^\prime_{\rm 5w} \right\}^{-1}_{NN} \right).
\end{equation}

From this result, we may set
\begin{eqnarray}
\label{eq:effective-Dirac-operator-I}
a D_{\rm eff}^{(N)} 
&=& 1- \frac{1}{a_5}
\left\{ \overline{D}^\prime_{\rm 5w} \right\}^{-1}_{NN} \\
&=&  1- \frac{1}{a_5}
\left(  P_R \left\{ \overline{D}_{\rm 5w} \right\}^{-1}_{NN} P_L 
       + P_L \left\{ \overline{D}_{\rm 5w} \right\}^{-1}_{11} P_R \right. 
\nonumber\\
&& \qquad \qquad 
\left.
       + P_R \left\{ \overline{D}_{\rm 5w} \right\}^{-1}_{N1} P_R
       + P_L \left\{ \overline{D}_{\rm 5w} \right\}^{-1}_{1N} P_L \right).
\end{eqnarray}
Thus the effective four-dimensional Dirac operator can be
expressed in terms of the inverse of the five-dimensional 
Wilson-Dirac operator {\em with the anti-periodic boundary condition}.
Since the gauge field is four-dimensional, 
the inverse of this five-dimensional Wilson-Dirac operator 
may be expressed as follows:
\begin{equation}
\left\{a_5  \overline{D}_{\rm 5w}  \right\}_{st}^{-1}  
= \frac{1}{N}\sum_p 
e^{i p(s-t)}
\left\{ i \gamma_5 \sin p+ 1-\cos p
        + a_5 \left(D_{\rm w}- \frac{m_0}{a}\right) \right\}^{-1}  , 
\end{equation}
where the summation is taken over the discrete 
momenta ${p=\frac{2\pi}{N}(k-\frac{1}{2})}$ $(k=1,2,\cdots,N)$ and
$D_{\rm w}$ is the four-dimensional Wilson-Dirac operator
\begin{equation}
D_{\rm w} 
= \sum_\mu 
\left\{\gamma_\mu \frac{1}{2}\left(\nabla_\mu+\nabla_\mu^\ast\right) 
-\frac{a}{2} \nabla_\mu\nabla_\mu^\ast \right\}. 
\end{equation}
Then the effective Dirac operator may be expressed as follows:
\begin{eqnarray}
\label{eq:effective-D-at-N-Integral-rep}
a D_{\rm eff}^{(N)} 
&=& 1 
- P_R \, 
\frac{1}{N}\sum_p
\frac{1}{i \gamma_5 \sin  p + 1-\cos p 
+ a_5 \left(D_{\rm w}- \frac{m_0}{a}\right) }
P_L  \nonumber\\
&& \quad
- P_L \,
\frac{1}{N}\sum_p
\frac{1}{i \gamma_5 \sin  p + 1-\cos p 
+ a_5 \left(D_{\rm w}- \frac{m_0}{a}\right) }
P_R  \nonumber\\
&& \quad
+ P_R \, 
\frac{1}{N}\sum_p
\frac{e^{-ip}}
{i \gamma_5 \sin p + 1-\cos p 
+ a_5 \left(D_{\rm w}- \frac{m_0}{a}\right) } 
P_R  \nonumber\\
&& \quad
+ P_L \,
\frac{1}{N}\sum_p
\frac{e^{+ip}}
{i \gamma_5 \sin p + 1-\cos p 
+ a_5 \left(D_{\rm w}- \frac{m_0}{a}\right) } 
P_L . \nonumber \\
\end{eqnarray}
In the limit $N \rightarrow \infty$, the summation over the 
discrete momentum 
reduces to the continuous integral:
\begin{equation}
\frac{1}{N}\sum_{p=\frac{2\pi}{N}(k-\frac{1}{2})}   
\Longrightarrow 
\int^\pi_{-\pi} \frac{dp}{2\pi} .
\end{equation}

Note that, since we do not use the transfer matrix in this
derivation, this expression could hold true even if $B$ is not positive
definite and the transfer matrix is not defined consistently.
$m_0$ can be chosen as any value within $ m_0 \in [0,2]$ (when
$a_5=a$), as long as the five-dimensional Wilson-Dirac operator
with the anti-periodic boundary condition in the fifth dimension 
is not singular and invertible.

\subsection{The Ginsparg-Wilson relation}
\label{subsec:GW-relation}

Next we discuss the Ginsparg-Wilson relation for the effective
Dirac operator in the integral representation.
As we have seen in the previous subsection, 
the effective Dirac operator, $D_{\rm eff}$, is defined by 
\begin{equation}
a D_{\rm eff} 
= 1- \frac{1}{a_5}
\left\{ \overline{D}^\prime_{\rm 5w} \right\}^{-1}_{NN} \qquad
(N= \infty).
\end{equation}
If it would satisfies the Ginsparg-Wilson relation 
\begin{equation}
\label{eq:GW-relation-effective}
\gamma_5 D_{\rm eff} + D_{\rm eff} \gamma_5 
= 2 a D_{\rm eff} \gamma_5 D_{\rm eff}, 
\end{equation}
then the following identity must hold true 
in the limit of $N\rightarrow \infty$:
\begin{equation}
\label{eq:GW-relation-prime}
\gamma_5 
\left\{ \bar D^\prime_{\rm 5w} \right\}^{-1}_{NN}
+ 
\left\{ \bar D^\prime_{\rm 5w} \right\}^{-1}_{NN} \gamma_5   
= 
\frac{2}{a_5}
\left\{ \bar D^\prime_{\rm 5w} \right\}^{-1}_{NN} 
\gamma_5 
\left\{ \bar D^\prime_{\rm 5w} \right\}^{-1}_{NN} \quad
(N= \infty).
\end{equation}

We may compare this identity with Eq.~(\ref{eq:chiral-property-PV-D})
which express the chiral property of 
$\bar D^\prime_{\rm 5w}$ under the chiral transformation introduced 
by Furman and Shamir Eq.~(\ref{eq:chiral-transformation-of-DW}).
The latter we may write 
\begin{eqnarray}
\label{eq:chiral-property-domain-wall-D-inverse}
\left\{ \Gamma_5 
\left\{ \overline{D}_{\rm 5w}^\prime \right\}^{-1} 
+ \left\{ \overline{D}_{\rm 5w}^\prime \right\}^{-1} \Gamma_5 \right\}_{st}
&=& \frac{2}{a_5} \, 
\left\{ \overline{D}_{\rm 5w}^\prime \right\}^{-1}_{s \frac{N}{2}} 
 \gamma_5 
\left\{ \overline{D}_{\rm 5w}^\prime \right\}^{-1}_{\frac{N}{2} t}
\nonumber\\
&& +\frac{2}{a_5} \, 
\left\{ \overline{D}_{\rm 5w}^\prime \right\}^{-1}_{s N} 
 \gamma_5 
\left\{ \overline{D}_{\rm 5w}^\prime \right\}^{-1}_{N t} . \nonumber\\
\end{eqnarray}
Setting $s=t=N$, we obtain 
\begin{eqnarray}
\gamma_5 
\left\{ \overline{D}_{\rm 5w}^\prime \right\}^{-1}_{NN} 
+ \left\{ \overline{D}_{\rm 5w}^\prime \right\}^{-1}_{NN} \gamma_5 
&=& \frac{2}{a_5} \, 
\left\{ \overline{D}_{\rm 5w}^\prime \right\}^{-1}_{N \frac{N}{2}} 
 \gamma_5 
\left\{ \overline{D}_{\rm 5w}^\prime \right\}^{-1}_{\frac{N}{2} N}
\nonumber\\
&& +\frac{2}{a_5} \, 
\left\{ \overline{D}_{\rm 5w}^\prime \right\}^{-1}_{N N} 
 \gamma_5 
\left\{ \overline{D}_{\rm 5w}^\prime \right\}^{-1}_{N N} . \nonumber\\
\end{eqnarray}
Then we can see that Eq.~(\ref{eq:GW-relation-prime})
is equivalent to the following condition in 
the limit of $N\rightarrow \infty$:
\begin{equation}
\label{eq:GW-relation-residue}
  \left\{ \overline{D}_{\rm 5w}^\prime \right\}^{-1}_{N \frac{N}{2}} 
  = 0  \qquad (N\rightarrow \infty).
\end{equation}

As we will see below, this condition is fulfilled
as long as the five-dimensional Wilson-Dirac operator 
with the anti-periodic boundary condition is not singular and
invertible. Then we obtain Eq.~(\ref{eq:GW-relation-prime}) and 
Eq.~(\ref{eq:GW-relation-effective}), the Ginsparg-Wilson relation 
for the effective Dirac operator.

\section{Positivity of the square of the five-dimensional 
Wilson-Dirac operator with anti-periodic boundary condition}
\label{sec:positivity}
\reseteqnum

From Eqs.~(\ref{eq:effective-D-at-N-Integral-rep}) and
its $N \rightarrow \infty$ limit,
we see that 
for the effective four-dimensional Dirac operator to be defined
consistently,
it is required that 
the five-dimensional Wilson-Dirac operator with the anti-periodic 
boundary condition should be non-singular and invertible. 
In this section, we examine the positivity of the five-dimensional
Wilson-Dirac operator square.

To examine this requirement, we evaluate the square of the 
five-dimensional Wilson-Dirac operator.
Setting $a_5=a$ for simplicity, we have 
\begin{eqnarray}
&&\left\{ i \gamma_5 \sin p+ 1-\cos p + \left(a D_{\rm w}-m_0 \right)
 \right\}^\dagger 
\left\{ i \gamma_5 \sin p+ 1-\cos p + \left(a D_{\rm w}-m_0 \right)
 \right\}  \nonumber\\
&& = 
4 \sin^2\left(p/2\right)
\left(1-m_0 -\frac{a^2}{2} \nabla_\mu\nabla_\mu^\ast\right) 
+ \left(a D_{\rm w}-m_0\right)^\dagger \left(a D_{\rm w}-m_0\right) .
\end{eqnarray}
For $m_0=1$, the first term is positive semi-definite and then 
the positivity of 
$a^2 \overline{D}_{\rm 5w}^\dagger\overline{D}_{\rm 5w}$ 
is entirely determined by the positivity of the four-dimensional
Wilson-Dirac operator square, $(a D_{\rm w}-1)^\dagger(a D_{\rm w}-1)$.
According to the result of \cite{locality-of-overlap-D}, 
if the plaquette variables $U(p)$ are uniformly bounded as
\begin{equation}
  \parallel 1- U(p) \parallel < \epsilon ,
\end{equation}
we obtain
\begin{eqnarray}
&&
\parallel
\left\{ i \gamma_5 \sin p+ 1-\cos p + \left(a D_{\rm w}-1\right)
 \right\}^\dagger 
\left\{ i \gamma_5 \sin p+ 1-\cos p + \left(a D_{\rm w}-1\right)
 \right\}  
\parallel
\nonumber\\
&& = 
\parallel
4 \sin^2\left(p/2\right)
\left( -\frac{a^2}{2} \nabla_\mu\nabla_\mu^\ast\right) 
+ \left(a D_{\rm w}-1\right)^\dagger \left(a D_{\rm w}-1\right) 
\parallel
\nonumber\\
&& \ge  1- 30 \epsilon .
\end{eqnarray}
For the generic value of $m_0 \in [0,2]$, we also 
obtain \cite{adams-at-chiral99,luscher-unpublished,bound-neuberger}
\begin{eqnarray}
&&
\parallel
\left\{ i \gamma_5 \sin p+ 1-\cos p + \left(a D_{\rm w}-m_0 \right)
 \right\}^\dagger 
\left\{ i \gamma_5 \sin p+ 1-\cos p + \left(a D_{\rm w}-m_0 \right)
 \right\} 
\parallel
 \nonumber\\
&& \ge  \left\{ (1- 30 \epsilon)^{\frac{1}{2}}-|1-m_0| \right\}^2 
\qquad {\rm if} \quad 1- 30 \epsilon > |1-m_0|^2 .
\end{eqnarray}
Recently, it has been shown by Neuberger \cite{bound-neuberger} that
the constant $30$ in the above bounds can be improved to $6(2+\sqrt{2})$.

From these considerations, 
we may assume the positive lower and upper bounds of 
the square of the five-dimensional Wilson-Dirac operator 
with the anti-periodic boundary condition as
\begin{equation}
  0 < \tilde \alpha \le \, 
\left\{ 
4 \sin^2\left(p/2\right)
B + \left(a D_{\rm w}-m_0\right)^\dagger
\left(a D_{\rm w}-m_0\right)
\right\}
\, 
\le \tilde \beta ,
\end{equation}
under the following condition,
\begin{equation}
\label{eq:condition-on-plaquett}
  \parallel 1- U(p) \parallel < \epsilon , \qquad
  \epsilon < \frac{1}{6(2+\sqrt{2})}\left(1 - |1-m_0|^2\right).
\end{equation}

\section{Expansion with Chebycheff polynomials and an exponential bound}
\label{sec:exponential-bound}
\reseteqnum

Given the bounds on the square of the five-dimensional Wilson-
Dirac operator with the anti-periodic boundary condition, 
we will derive in this section
exponential bounds on the effective four-dimensional 
Dirac operator and its differentiations with respect to the gauge field. 
We will also obtain an exponential bound on the inverse of 
the five-dimensional Wilson-Dirac operator which is needed to prove 
Eq.~(\ref{eq:GW-relation-residue}) and the Ginsparg-Wilson relation. 

\subsection{Locality bounds}
\label{subsec:locality-bound}

From Eqs.~(\ref{eq:effective-D-at-N-Integral-rep}) and
its $N \rightarrow \infty$ limit,
we see that 
the locality property of the effective Dirac operator of the domain-wall
fermion is determined by the locality properties of the following
operators in the integral representation:
\begin{equation}
\label{eq:sum-in-effective-D}
I^{(N)} =
\frac{1}{N} \sum_p 
\left\{1, e^{+ip}, e^{-ip} \right\}
\frac{1}
{i \gamma_5 \sin p + 1-\cos p + \left(a D_{\rm w}-m_0\right) }   
\end{equation}
and
\begin{equation}
\label{eq:integral-in-effective-D}
I =\int^\pi_{-\pi}\frac{dp}{2\pi} 
\left\{1, e^{+ip}, e^{-ip} \right\}
\frac{1}
{i \gamma_5 \sin p + 1-\cos p + \left(a D_{\rm w}-m_0\right) }   .
\end{equation}
The integrand can be written as 
\begin{eqnarray}
&& \frac{1}
{i \gamma_5 \sin p + 1-\cos p + \left(a D_{\rm w}-m_0\right) }  
\left\{1, e^{+ip}, e^{-ip} \right\}
\nonumber\\
&& 
= 
\frac{1} 
     {
4 \sin^2\left(p/2\right) B 
+ \left(a D_{\rm w}-m_0\right)^\dagger
\left(a D_{\rm w}-m_0\right)
} \times  \nonumber\\
&& \qquad\qquad\qquad
\left( 1 - P_R e^{ip} - P_L e^{-ip} + 
\left(a D_{\rm w}-m_0\right)^\dagger\right) 
\left\{1, e^{+ip}, e^{-ip} \right\} .
\nonumber\\
\end{eqnarray}
From this expression it is clear that
the operators in the numerator are local and bounded.
Then we may omit these operators in the following considerations. 

We can obtain an expansion of the integrand
using the generating function of the Chebycheff 
polynomials \cite{bunk}
\begin{equation}
  \frac{1}{1- 2 t z + t^2 } 
= \sum_{k=0}^\infty t^k U_k(z) , \qquad 
\|  U_k(z) \| \, \le \, U_k(1)= k .
\end{equation}
Following \cite{locality-of-overlap-D}, we set 
\begin{equation}
t= e^{-\tilde \theta}, \qquad \cosh 
\theta = \frac{\tilde \beta + \tilde \alpha}
              {\tilde \beta - \tilde \alpha} ,
\end{equation}
and
\begin{equation}
  z= \frac{\tilde \beta+\tilde \alpha 
- 2 
\left\{ 
4 \sin^2\left(p/2\right) B 
+ \left(a D_{\rm w}-m_0\right)^\dagger
\left(a D_{\rm w}-m_0\right)
\right\}
            }
          {\tilde \beta-\tilde \alpha}.
\end{equation}
Then we obtain
\begin{equation}
\label{eq:series-expansion-with-Chebycheff-polynomials}
\frac{1}{
4 \sin^2\left(p/2\right) B 
+ \left(a D_{\rm w}-m_0\right)^\dagger
\left(a D_{\rm w}-m_0\right)
}
= \frac{4 t}{\tilde \beta-\tilde \alpha}\sum_{k=0} t^k U_k( z ) ,
\end{equation}

This defines an expansion in terms of the square of the Wilson-Dirac 
operator and $B$ with only 
nearest-neighbor and next-to-nearest-neighbor couplings.
In order to contribute to the kernel of the operator 
Eq.~(\ref{eq:series-expansion-with-Chebycheff-polynomials})
between two lattice sites $x$ and $y$ of 
the lattice distance $d(x,y)=\vert x-y \vert$, 
the order of the polynomials $U_k(x)$ in the expansion
must be greater than $\frac{d(x,y)}{2a}$: 
\begin{equation}
    k \ge \frac{d(x,y)}{2a} 
\end{equation}
Then for the given distance $d(x,y)$, 
the series expansion 
Eq.~(\ref{eq:series-expansion-with-Chebycheff-polynomials}) 
can be arranged as follows:
\begin{eqnarray}
&&
\frac{4 t}{\tilde \beta-\tilde \alpha}
\exp \left\{ - \frac{\tilde \theta}{2a}d(x,y)\right\} .
\sum_{k=0} t^k U_{k+ d/2a}( z ) .
\end{eqnarray}
Noting the bound on the polynomials, $\| U_k(z) \| \le k$, we obtain
\begin{eqnarray}
&& 
\left\|
\frac{1}{
4 \sin^2\left(p/2\right) B 
+ \left(a D_{\rm w}-m_0\right)^\dagger
\left(a D_{\rm w}-m_0\right)
}(x,y)
\right\|
\nonumber\\
&& \qquad \qquad 
\le \frac{4 t}{\tilde \beta-\tilde \alpha}
\exp \left\{ - \frac{\tilde \theta}{2a}d(x,y)\right\} .
\sum_{k=0} t^k \| U_{k+ d/2a}( z )\| 
\nonumber\\
&& \qquad \qquad \le \, 
\frac{4 t}{\tilde \beta-\tilde \alpha} 
\left( 
\frac{1}{1-t} \frac{d(x,y)}{2a} + \frac{t}{(1-t)^2} 
\right)
  \exp \left\{ - \frac{\tilde \theta}{2a}d(x,y)\right\} . 
\nonumber\\
\end{eqnarray}
Since the summation over the momentum in $I^{(N)}$
(the integration in $I$) is normalized to unity, 
the above bounds implies the exponential bound for the integrals, 
$I^{(N)}$ and $I$.

As for the differentiations of the effective Dirac operator, 
we can also derive the exponential bounds, following 
\cite{locality-of-overlap-D}.
We consider the differentiations of the Chebycheff expansion, 
Eq.~(\ref{eq:series-expansion-with-Chebycheff-polynomials}).
We first introduce an integral representation for the Chebycheff 
polynomials:
\begin{equation}
  U_k(z) = \oint \frac{d\omega}{2\pi} \omega^{-k-1} 
\frac{1}{\omega^2 - 2\omega z + 1 }, 
\end{equation}
where the integration is defined along a circle in the complex
plane centered at the origin. The radius $r$ of the circle
should be strictly less than $1$ to avoid the singularities
of the integrand.
The denominator of the integrand can be factorized according to
\begin{equation}
\omega^2 - 2\omega z + 1 
= (\omega - u^\dagger) ( \omega - u ), \qquad
u =z +i (1-r)^{1/2} .
\end{equation}
Since, $u^\dagger u = 1$, it is clear that
\begin{equation}
\parallel 
\left(\omega^2 - 2\omega z + 1 \right)^{-1} 
\parallel 
\le \left( 1 - r \right)^{-2} .
\end{equation}

If we denote the differentiation of $z$ with respect to
the gauge fields as $\dot z$, then we obtain 
\begin{equation}
  \parallel \dot {U}_k(z) \parallel 
\le 2 \parallel \dot{z} \parallel r^{-k} \left(1-r\right)^{-4}.
\end{equation}
We may now adjust the radius $r$ so that the factor 
$r^{-k} \left(1-r\right)^{-4}$ is minimized. We obtain
\begin{equation}
\parallel \dot {U}_k(z) \parallel \le  {\rm constant} \ 
\parallel \dot{z} \parallel  (1+k)^4. 
\end{equation}
With these bounds, we can see that the differentiated series 
Eq.~(\ref{eq:series-expansion-with-Chebycheff-polynomials})
is also exponentially convergent with the same exponent as the 
original series. By similar estimations, we can see that this is also true 
for higher-order differentiations
(each differentiation give rise to an additional
factor of $(1+k)^2$ in the bound on the Chebycheff polynomials).

\subsection{Exponential bounds in the fifth-direction and \\
the Ginsparg-Wilson relation}
\label{subsec:bound-in-fifth-dim}

We next consider the exponential bound on the inverse of 
the five-dimensional Wilson-Dirac operator which is necessary to prove 
Eq.~(\ref{eq:GW-relation-residue}) and the Ginsparg-Wilson relation. 
From the above derivation of the exponential bounds for
the summation and integral Eqs.~(\ref{eq:sum-in-effective-D}) 
and (\ref{eq:integral-in-effective-D}), 
we can see that the same bound holds true for the inverse of
the five-dimensional Wilson-Dirac operator, $\overline{D}_{\rm 5w}$,
itself:
\begin{equation}
\label{eq:inverse-5dim-Wilson-D}
\left\{a \overline{D}_{\rm 5w} \right\}^{-1}_{st}=
\frac{1}{N} \sum_p 
\frac{e^{+ip(s-t)}}
{i \gamma_5 \sin p + 1-\cos p + \left(a D_{\rm w}-m_0\right) }   
\end{equation}
In fact, we can obtain 
\begin{equation}
\left\| \left\{a^2 \overline{D}_{\rm 5w}^\dagger  
\overline{D}_{\rm 5w}\right\}^{-1}(x,s;y,t) 
\right\| \le  \, 
C \, 
\exp \left\{ - \frac{\tilde \theta}{2a}d_5(x,s;y,t)\right\} ,
\end{equation}
where $d_5(x,s;y,t)=|x-y|+{\rm min}(|s-t|,N-|s-t|)$ and 
\begin{equation}
C=\frac{4 t}{\tilde \beta-\tilde \alpha} 
\left( 
\frac{1}{1-t} \frac{d_5(x,s;y,t)}{2a} + \frac{t}{(1-t)^2} 
\right).
\end{equation}

From this bound, it follows immediately that 
\begin{equation}
\label{eq:GW-relation-residue-vanishes}
\lim_{N\rightarrow \infty}
  \left\{ \overline{D}_{\rm 5w}^\prime \right\}^{-1}_{N \frac{N}{2}} 
  = 0  .
\end{equation}
This completes the proof of the Ginsparg-Wilson relation under
the condition on the plaquette
variables Eq.~(\ref{eq:condition-on-plaquett}). 
Note again that this proof
does not refer to the transfer matrix and it applies for any value
of $m_0 \in [0,2]$ .

\subsection{Singular case}
\label{subsec:singular-case}

In this subsection, we examine locality of the effective Dirac operator
Eq.~(\ref{eq:effective-Dirac-operator})
with the singular gauge configuration for which 
isolated eigenvalues of the hermitian Wilson-Dirac operator collapse 
to zero:
\begin{equation}
  H \phi_0(x) = \lambda \phi_0(x), \qquad \lambda \simeq 0 ,
\end{equation}
where
\begin{equation}
H=\gamma_5 ( D_{\rm w} - \frac{m_0}{a} ) .
\end{equation}
We will argue that the effective Dirac operator remains local 
even with such singular gauge configurations.

For this purpose, however, the integral representation 
and the Chebycheff expansion in terms of the five-dimensional 
Wilson-Dirac operator, considered so far, does not seem to be useful. 
When the isolated near-zero mode occurs 
in the four-dimensional hermitian Wilson-Dirac operator, 
it is associated with many modes with small fifth momenta 
in the spectrum of the five-dimensional Wilson-Dirac
operator (with the anti-periodic boundary condition).
This means that the continuum spectrum would collapse to zero 
in the five-dimensional Wilson-Dirac operator in the limit 
$N\rightarrow \infty$.
Then the separation of the effect of the near-zero mode
does not seem easy in this representation (cf. \cite{locality-of-overlap-D}). 
Therefore, in this section, we use the formula for the effective 
action in terms of the transfer matrix and $\widetilde H$.

In \cite{locality-of-overlap-D}, it has been proved rigorously that
the contribution of the near-zero-mode 
to Neuberger's Dirac operator,
\begin{equation}
\left.  \frac{H}{\sqrt{H^2}} \right|_{\rm near-zero} , 
\end{equation}
remains local. This result can be understood from the localization 
properties of the eigenvectors of the near-zero modes.
In fact, it is well localized with exponentially
decaying tails \cite{locality-of-overlap-D,
Edwards-et-al-small-eigenvalues-localization}.

Since $\widetilde H$ is related to  $H$ by 
the formula 
\begin{equation}
e^{ - a_5 \tilde H } + e^{ + a_5 \tilde H }  -2 
= a_5^2 H \frac{1}{B} H , 
\end{equation}
the exact zero mode of $\widetilde H$ is the exact zero mode of 
$H$ \cite{overlap,
Edwards-et-al-small-eigenvalues-localization,
shamir-modified-PF, luscher-private}. 
Therefore, the contribution of the zero mode 
of $\widetilde H$ to the effective Dirac operator 
is identical to the contribution of the zero mode 
of $H$ to Neuberger's Dirac operator,
\begin{equation}
\left.  \frac{\widetilde H}{\sqrt{\widetilde H^2}} \right|_{\rm zero} 
=  \left. \frac{H}{\sqrt{H^2}} \right|_{\rm zero} 
= \lim_{\lambda \rightarrow 0} 
{\rm sign}(\lambda) \, \phi_0(x) \phi_0^\dagger(y)  ,
\end{equation}
and it remains local, according to the result of \cite{locality-of-overlap-D}.
It is expected that this localization 
properties persist also for the contribution of the 
near-zero modes of $\widetilde H$.

It is desirable to make the above argument rigorous.
We will leave this issue for future study.

\section{Discussion}
\label{sec:summary-discussion}
\reseteqnum

We have argued locality in the domain-wall fermion approach
through its effective four-dimensional Dirac operator.
As expected, all the properties proved rigorously for Neuberger's 
Dirac operator holds true for the effective Dirac operator.
In particular, we have shown explicitly that 
the locality properties of the domain-wall fermion
depends crucially on the spectrum of the four-dimensional 
Wilson-Dirac operator, which is closely related to that of 
the five-dimensional Wilson-Dirac operator (with the anti-periodic 
boundary condition). Then we can see that the bound for the plaquette 
variables leads to the locality bound for the effective Dirac operator.
We have also shown that
the effective Dirac operator satisfies the Ginsparg-Wilson relation 
with the same bound for plaquette variables.

The approach to the chiral symmetry limit from a finite $N$ 
is the most important issue for the practical implementation 
of exact chiral symmetry using the domain-wall fermion 
\cite{DWF-columbia,DWF-tsukuba-fine-coarse,DWF-RBC}
\cite{shamir-modified-PF,DWF-strong-coupling-expansion,better-DWF}.
In order to examine the effect of the finite $N$,
the explicit breaking term in the axial Ward-Takahashi 
identity has been measured, among other physical quantities.
This breaking term can be written by the correlation function
between the middle and 
the boundary of the fifth dimension \cite{kikukawa-noguchi}:
\begin{equation}
\left\{ a_5 D_{\rm 5w}^\prime \right\}^{-1}_{\frac{N}{2},N}
=
\frac{1}{2 \cosh \frac{N}{2}a_5 \widetilde H } 
\times \left(\frac{1}{a} { D_{\rm eff}^{(N)}}^{-1} \right) (x,y) .
\end{equation}
From the point of view of the effective four-dimensional action, 
this effect may be examined through the breaking term in the
Ginsparg-Wilson relation.\footnote{
Numerical study of the Ginsparg-Wilson relation of 
the effective Dirac operator is found in \cite{borici2}.}
As we have seen in the 
section~\ref{subsec:bound-in-fifth-dim}, 
it is given by the similar correlation function 
defined through the five-dimensional Wilson-Dirac operator
with the anti-periodic boundary condition:
\begin{eqnarray}
\left\{ a_5 \overline{D}_{\rm 5w}^\prime \right\}^{-1}_{\frac{N}{2},N}
&=& \frac{1}{2 \cosh \frac{N}{2}a_5 \widetilde H } 
\nonumber\\
&=& \frac{1}{N}\sum_p 
\frac{ ( e^{-i p } )^\frac{N}{2} }
{i \gamma_5 \sin p+ 1-\cos p
        + a_5 \left(D_{\rm w}- \frac{m_0}{a}\right) } .
\end{eqnarray}
It is the smallest eigenvalue of square of the four-dimensional 
Wilson-Dirac operator which determines the behavior of the 
breaking term in the limit $N\rightarrow \infty$.
When the isolated near-zero mode occurs 
in the four-dimensional hermitian Wilson-Dirac operator, 
it is associated with many modes with small fifth momenta 
in the spectrum of the five-dimensional Wilson-Dirac
operator (with the anti-periodic boundary condition).
This means that the continuum spectrum tends to collapse to zero and
the lower bound for this continuum spectrum determines the rate of
the exponential decay in the limit $N\rightarrow \infty$. 
Therefore, it would be important 
to examine the behavior of the near-zero 
modes as done in \cite{locality-of-overlap-D},
also in the context of the domain-wall fermion,
as suggested in \cite{DWF-tsukuba-fine-coarse}.
It is also desirable to clarify the nature of the distribution 
of small eigenvalues of the four-dimensional Wilson-Dirac operator 
with a negative mass, for the gauge field configurations 
used in the current 
simulations \cite{Edwards-et-al-small-eigenvalues-distribution}.

\section*{Acknowledgments}

The intensive discussions at the Summer Institute 99 at Yamanashi, Japan,
were very suggestive and useful to complete this work. 
The author would like to thank S.~Aoki, T.~Onogi, Y.~Taniguchi, T.~Izubuchi, 
K.~Nagai, J.~Noaki, K.~Nagao, N.~Ukita, and H.~So for enlightening 
discussions. He is grateful to D.A.~Adams and H.~Neuberger
for the correspondences concerning the bound of the hermitian 
Wilson-Dirac operator. The author would like to thank T.-W.~Chiu 
for the kind hospitality at Chiral '99 in Taipei. 
The author is also grateful to M.~L\"uscher and A.~Bori\c ci for 
the comments on the first version of this article.
This work is supported in part by Grant-in-Aid for Scientific 
Research of Ministry of Education (\#10740116).


\newpage

\begin{thebibliography}{99}

\bibitem{overlap-D}
H.~Neuberger, \plb{417}{1998}{141}; Phys. Lett. {\bf B427} (1998) 353.

\bibitem{overlap}
R.~Narayanan and H.~Neuberger, 
\npb{412}{1994}{574};
\prl{71}{1993}{3251};
\npb{443}{1995}{305}.

\bibitem{ginsparg-wilson-rel}
P.~H.~Ginsparg, K. G. Wilson, \prd{25}{1982}{2649}. 

\bibitem{locality-of-overlap-D}
P.~Hern\'andes, K.~Jansen, M.~L\"uscher, 
Nucl.Phys. B552 (1999) 363-378, {\tt hep-lat/9808010}.

\bibitem{index-theorem-at-finite-lattice}
P.~Hasenfratz, V.~Laliena, F.~Niedermayer, Phys. Lett. B427 (1998) 125.

\bibitem{chiral-anomaly-abelian}
M.~L\"uscher, Nucl. Phys. B538 (1999) 515.

\bibitem{abelian-chiral}
M.~L\"uscher, Nucl. Phys. B549 (1999) 295.

\bibitem{nonabelian-chiral}
M.~L\"uscher, ``Weyl fermions on the lattice and the non-abelian
gauge anomaly'', {\tt hep-lat/9904009}.

\bibitem{fujiwara-suzuki-wu}
T.~Fujiwara, H.~Suzuki, K.~Wu, 
``Non-commutative Differential Calculus and the Axial Anomaly 
in Abelian Lattice Gauge Theories'', {\tt hep-lat/9906015};
Phys. Lett. B463 (1999) 63.

\bibitem{adams-6d}
D.H.~Adams, ``A topological aspect of the non-abelian anomaly 
for Weyl fermions on the lattice'', {\tt hep-lat/9910036}.

\bibitem{lat99-luscher}
M.~L\"uscher,
plenary talk 
at the International Symposium on Lattice Field Theory, Pisa, Italy,
June 29-July 3, 1999.

\bibitem{borici}
A.~Bori\c ci, Phys. Lett. B453 (1999) 46; {\tt hep-lat/9910045}.

\bibitem{domain-wall-fermion}
D.~B.~Kaplan, \plb{288}{1992}{342}. 

\bibitem{boundary-fermion}
Y.~Shamir, \npb{406}{1993}{90}. 

\bibitem{boundary-fermion-QCD}
V.~Furman, Y.~Shamir, \npb{439}{1995}{54}.

\bibitem{blum-soni}
T.~Blum, A.~Soni, Phys. Rev. D56 (1997) 174;
Phys. Rev. Lett. 79 (1997)  3595.

\bibitem{blum-soni-wingate}
T.~Blum, A.~Soni, M.~Wingate, Phys. Rev. D60 (1999) 114507.

\bibitem{vranas-pauli-villars}
P.~Vranas, Phys. Rev. D57 (1998) 1415.

\bibitem{DWF-columbia}
P. Chen {\it et al.} (Columbia Univ. Collaboration), 
Phys. Rev. D59 (1999) 054508; 
Nucl. Phys. Proc. Suppl. 73 (1999) 204;
Nucl. Phys. Proc. Suppl. 73 (1999) 207;
Nucl. Phys. Proc. Suppl. 73 (1999) 456;
{\tt hep-lat/9812011};
{\tt hep-lat/9903024};
{\tt hep-lat/9909140};
{\tt hep-lat/9911002}.

\bibitem{DWF-tsukuba-oneloop-renormalization-factor}
S. Aoki, Y. Taniguchi, Phys. Rev. D59 (1999) 054510;
Nucl. Phys. Proc. Suppl. 63 (1998) 290;
S.~Aoki, T.~Izubuchi, Y.~Kuramashi, Y.~Taniguchi, 
Phys. Rev. D59 (1999) 094505; Phys. Rev. D60 (1999) 114504.
S.~Aoki, Y.~Taniguchi, Phys. Rev. D59 (1999) 094506;
S.~Aoki, T.~Izubuchi, J.~Noaki, Y.~Kuramashi, Y.~Taniguchi, 
{\tt hep-lat/9909049}.

\bibitem{DWF-tsukuba-scaling}
J.~Noaki, Y.~Taniguchi, 
``Scaling property of domain-wall QCD in perturbation theory''
{\tt  hep-lat/9906030}.

\bibitem{DWF-tsukuba-fine-coarse}
S.~Aoki, T.~Izubuchi, Y.~Kuramashi, Y.~Taniguchi, 
{\tt hep-lat/9909154};
A.~Ali Khan {\it et al.} (CP-PACS Collaboration), {\tt hep-lat/9909049}.

\bibitem{epsilon-prime-over-epsilon}
T. Blum {\it et al.} (RIKEN-BNL-Columbia Collaboration), 
``A first study of epsilon'/epsilon on the lattice using 
domain wall fermions'', {\tt hep-lat/9908025}.

\bibitem{DWF-RBC}
T.~Blum~{\it~et~al.} (RIKEN-BNL-Columbia~Collaboration), 
{\tt hep-lat/9909093;hep-lat/9909101;hep-lat/9909117;hep-lat/9909382}.

\bibitem{chiu-DWF}
T.-W.~Chiu, {\tt hep-lat/9912005}.

\bibitem{truncated-overlap}
H.~Neuberger, Phys. Rev. {\bf D57} (1998) 5417.

\bibitem{kikukawa-noguchi}
Y.~Kikukawa, T.~Noguchi, {\tt hep-lat/9902022}.

\bibitem{exact-chiral-symmetry}
M. L\"uscher, Phys. Lett. {\bf B428} (1998) 342.

\bibitem{kikukawa-yamada}
Y.~Kikukawa, A.~Yamada, Phys. Lett. B448 (1999) 265.

\bibitem{adams}
D.A.~Adams, {\tt hep-lat/9812019}.

\bibitem{suzuki}
H.~Suzuki, Prog. Theor. Phys. 102 (1999) 141.

\bibitem{fujikawa}
K.~Fujikawa, Nucl. Phys. B546 (1999) 480.

\bibitem{chiu}
T.-W.~Chiu, Phys. Lett. B445 (1999) 371;
T.-W.~Chiu, T.-H.~Hsieh, hep-lat/9901011.

\bibitem{luscher-private} 
M.~L\"uscher, private communication.

\bibitem{rational-approximation}
H.~Neuberger, Phys. Rev. Lett. 81, (1998) 4060;
Int. J. Mod. Phys. C10 (1999) 1051;
Phys. Rev. D60 (1999) 065006;
{\tt hep-lat/9909043}.

\bibitem{borici2}
A.~Bori\c ci, {\tt hep-lat/9909057}; {\tt hep-lat/9912040}.

\bibitem{bound-neuberger}
H.~Neuberger, {\tt hep-lat/9911004}

\bibitem{bunk}
B.~Bunk, {\tt hep-lat/9805030}.

\bibitem{Edwards-et-al-small-eigenvalues-localization}
R.G.~Edwards, U.M.~Heller, R.~Narayanan, 
Nucl. Phys. B535 (1998) 403. 

\bibitem{Edwards-et-al-small-eigenvalues-distribution}
R.G.~Edwards, U.M.~Heller, R.~Narayanan, 
Phys. Rev. D60 (1999) 034502.

\bibitem{shamir-modified-PF}
Y.~Shamir, Phys. Rev. D59 (1999) 054506.

\bibitem{DWF-strong-coupling-expansion}
R.C.~Brower, B.~Svetitsky, {\tt hep-lat/9912019}.

\bibitem{better-DWF}
Y.~Shamir, {\tt hep-lat/9912027}.

\bibitem{adams-at-chiral99}
D.H.~Adams, contribution to Chiral '99, ``Workshop on Chiral
Gauge Theories'', Sept. 13-18, 1999, Taipei. {\tt hep-lat/0001014}.

\bibitem{luscher-unpublished}
M. L\"uscher, unpublished note. 



\end{thebibliography}
\end{document}